\documentclass[twocolumn,showpacs,amsmath,amssymb,appendix,prd,superscriptaddress]{revtex4}

\usepackage{graphicx}
\usepackage{dcolumn}
\usepackage{bm}
\usepackage{epsfig}
\usepackage{amsmath}
\usepackage{color}

\newcommand{\be}{\begin{equation}}
\newcommand{\ee}{\end{equation}}
\newcommand{\ba}{\begin{eqnarray}}
\newcommand{\ea}{\end{eqnarray}}
\def\eps{\epsilon}

\def\g{\gamma}
\def\G{\Gamma}

\begin{document}
\input{epsf}

\title{PeV--EeV neutrinos from GRB blastwave in IceCube and future
  neutrino telescopes}

\author{Soebur Razzaque } \email{srazzaque@uj.ac.za} \affiliation{
  Department of Physics, University of Johannesburg, PO Box 524,
  Auckland Park 2006, South Africa}

\author{Lili Yang} \email{lili.yang@ung.si} \affiliation{ Department
  of Physics, University of Johannesburg, PO Box 524, Auckland Park
  2006, South Africa} \affiliation{Laboratory for Astroparticle
  Physics, University of Nova Gorica, Vipavska 13, 5000 Nova Gorica,
  Slovenia}

\begin{abstract}
  Ultrahigh-energy cosmic rays (UHECRs), if accelerated in the
  gamma-ray burst (GRB) blastwave, are expected to produce PeV--EeV
  neutrinos by interacting with long-lived GRB afterglow photons.
  Detailed spectral and temporal properties of the flux of these
  neutrinos depend on the GRB blastwave evolution scenario, but can
  last for days to years time scale in contrast to the seconds to
  minutes time scale for ``burst'' neutrino flux contemporaneous with
  the prompt gamma-ray emission and which has been constrained by
  IceCube in the $\sim$ 50 TeV--2 PeV range. We compute expected
  neutrino events in IceCube in the PeV--EeV range from the blastwave
  of long-duration GRBs, both for the diffuse flux and for individual
  GRBs in the nearby universe. We show that IceCube will be able to
  detect the diffuse GRB blastwave neutrino flux after 5 years of
  operation, and will be able to distinguish it from the cosmogenic
  neutrino flux arising from GZK process in case the UHECRs are heavy
  nuclei. We also show that EeV neutrinos from the blastwave of an
  individual GRB can be detected with long-term monitoring by a future
  high-energy extension of IceCube for redshift up to $z\sim 0.5$.
\end{abstract}


\date{\today}
\maketitle

\section{Introduction}

The long-duration gamma-ray bursts are thought to be the sources of
ultrahigh-energy cosmic rays ($\gtrsim 10^{18}$ eV) in nature
\cite{Waxman:1995vg, Vietri:1995hs}. Acceleration of protons and/or
ions to ultrahigh energies can take place in the internal shocks
and/or external forward and reverse shocks. In cases of internal and
external reverse shocks, GRB jets must contain protons and/or ions in
the form of ejected material from the GRB central engine
\cite{Wang:2007xj}. In case of external forward shock, which is
responsible for GRB afterglow emission \cite{Meszaros:1993tv}, protons
and/or ions from circumburst medium, forming a blastwave, are
accelerated.  Interactions of ultrahigh-energy particles (assumed
dominated by protons) with the ambient prompt or afterglow photons via
$p\gamma\to \pi^\pm,\, K^\pm$ processes are expected to produce $\nu$
from decays such as $\pi^+/K^+ \to \mu^+ \nu_\mu \to e^+ \nu_e {\bar
  \nu}_\mu \nu_\mu$ \footnote{Note that proton-proton ($pp$)
  interactions can be dominating in the GRB jet in its very early
  stage of evolution, while the jet is still burrowing through the
  stellar envelope and the density of material in the jet is very high
  \cite{Razzaque:2003uv}.  During the time of prompt and afterglow
  emissions, the density of material is very low because of increased
  size of the jet and $pp$ interactions become less important.}.
Detection of these $\nu$ can identify GRBs as accelerators of UHECRs.

IceCube Neutrino Observatory has recently imposed constraining limits
on the GRB ``burst'' $\nu$ flux \cite{Waxman:1997ti, Guetta:2003wi,
  Dermer:2003zv, Murase:2005hy, Hummer:2011ms} in the $\sim 50$ TeV--2
PeV range \cite{Abbasi:2012zw}. The ANTARES neutrino telescope has
also imposed somewhat weaker limits on the same flux
\cite{Adrian-Martinez:2013sga}. The most recent IceCube GRB analysis
\cite{Aartsen:2014aqy} also constrains neutrino flux from a model for
prompt $\gamma$-ray emission alternate to the internal shocks model
\cite{Zhang:2012qy}. Such limits are extremely useful to understand
the content of the GRB jet, its velocity and distance of the $\g$-ray
emission region from the central engine. Non-detection of $\nu$ from a
recent, bright burst GRB 130427A \cite{Zhu:2013ufa} at relatively low
redshift ($z=0.34$) led to interesting constraints on the emission
radius and the bulk Lorentz factor ($\Gamma$) of the GRB jet
\cite{Gao:2013fra}. Detection of $\g$ rays, with energy exceeding 10
GeV, from many GRBs by the Large Area Telescope (LAT) onboard the {\it
  Fermi Gamma Ray Space Telescope} also implies that $\G \gg 100$
(see, e.g., Ref.~\cite{Gehrels:2013xd, Ackermann:2013zfa}), to avoid
{\it in-situ} $\g\g\to e^\pm$ pair production.  The fraction of
Fermi-LAT GRBs is of the order of $10\%$ of the rate of GRBs detected
by the gamma-ray burst monitor (GBM) onboard Fermi
\cite{Ackermann:2013zfa}.

Ultrahigh-energy proton interactions ($p\gamma$) in the GRB blastwave
can produce PeV--EeV neutrinos \cite{Dermer:2000yd, Li:2002dw,
  Razzaque:2013dsa, Xiao:2014vga}.  These $\nu$ can be produced
efficiently and for a much longer time scale, if the blastwave has a
high initial bulk Lorentz factor \cite{Razzaque:2013dsa}, as compared
to the PeV--EeV $\nu$ from reverse shock \cite{Waxman:1999ai,
  Dai:2000dj, Murase:2007yt} which may be absent if the GRB jet is
magnetic energy dominated.  In case GRBs are not the significant
sources of observed UHECRs, these $\nu$ can still probe baryon
loading, magnetic field, particle acceleration, etc.\ in the GRB jet.
Here we calculate expected $\nu$ flux from GRB blastwave and the event
rates in IceCube using diffuse flux from all GRBs in the history of
the universe. We compare these rates with those from the cosmogenic
$\nu$ flux arising from GZK processes. We also calculate $\nu$ events
from individual GRB blastwaves in nearby universe ($z\lesssim 0.5$)
and the rate of detecting GRBs in $\nu$ by IceCube.  Finally, we
estimate event rates for the GRB blastwave $\nu$ fluxes by the
proposed high-energy upgrade of IceCube, which we refer to as IceHEX,
that will increase effective area in the $\gtrsim 1$ PeV range by up
to two orders of magnitude \cite{icehex}.

We describe our $\nu$ flux model calculations in Sec.\ II and
calculate $\nu$ detection rates by IceCube and IceHEX in Sec\ III.  We
discuss our results and draw conclusions in Sec.\ IV.

\section{Neutrino flux models}

\subsection{Flux from individual bursts}

We adopt the neutrino flux model from an adiabatic blastwave in
constant density interstellar medium described in
Ref.~\cite{Razzaque:2013dsa}.  Cosmic rays (assumed protons) are
accelerated in the forward shock in this scenario to a maximum energy
of
\be
E_{p,s} = 2.3\times 10^{20} \,
\frac{n_0^{1/8}\eps_{B,-1}^{1/2} E_{55}^{3/8}}
{\phi (1+z)^{7/8} t_2^{1/8}} ~{\rm eV},
\label{Emax_ad_i}
\ee
where the isotropic-equivalent kinetic energy of the blastwave is $E_k
= 10^{55}E_{55}$ erg, $n_0$ is the number density of particles per
cubic centimeter in the circumburst medium, the fraction of shock
energy converting to magnetic field energy is $\epsilon_B =
0.1\eps_{B,-1}$ and $\phi^{-1} \lesssim 1$ is an efficiency factor for
proton acceleration. This energy is evaluated at a time $t=10^2t_2$ s
after the GRB explosion and is valid for a time after the blastwave
deceleration time scale
\be
t_{dec} =
33.3\, 
\frac{(1+z) E_{55}^{1/3}}{n_0^{1/3} \Gamma_{2.5}^{8/3}}  ~{\rm s},
\label{dec_time_ism}
\ee
where the initial bulk Lorentz factor of the blastwave, before
deceleration, is assumed $\Gamma_0 = 10^{2.5}\Gamma_{2.5}$. The flux
of this cosmic-ray protons (if escape freely from the blastwave) can
be calculated, assuming an $E_p^{-2}$ spectrum arising from a
shock-acceleration process, as
\ba 
E_p^2 J_p(E_p) &=& 4.8\times 10^{-9} \,
\frac{(1+z)^{1/4}\eps_p n_0^{1/4} E_{55}^{3/4}}
{\xi_1 t_2^{1/4} d_{28}^{2}} \cr
&& {\rm GeV~cm}^{-2}~{\rm s}^{-1},
\label{Jp_ad_i}
\ea
where $\eps_p \lesssim 1$ is the fraction of blastwave kinetic energy
carried by the shock-accelerated protons, $\xi =10\xi_1$ is a spectral
correction factor and a luminosity distance of $10^{28}d_{28}$ cm was
assumed for the GRB. The total energy in cosmic rays is ${\cal E}_{CR}
= \eps_p E_k/2$. Note that we have assumed $\eps_p$ to be a free
parameter in our calculation, as there is no observational evidence
yet to constrain it. Neutrino detection from GRBs in principle can be
used to put constraint on $\eps_p$.

The opacity for $p\gamma$ interactions of these protons with
forward-shock afterglow synchrotron photons can be estimated as
\be
\tau_{p\g} (E_{p,l}) = 0.7\, 
\frac{\eps_{B,-1}^{1/2}n_0 E_{55}^{1/2} t_{2}^{1/2}}
{(1+z)^{1/2}},
\label{opt_ad_i}
\ee
at an energy
\be
E_{p,l} =  
\frac{1.3\times 10^{8}\, t_2^{3/4}}
{(1+z)^{7/4} \eps_{B,-1}^{1/2} \eps_{e,-1}^{2} 
n_0^{1/4} E_{55}^{1/4}} ~{\rm GeV},
\label{CRbreaks_ad_i_lo}
\ee
which corresponds to the break energy in the afterglow synchrotron
spectrum due to the characteristic synchrotron photon energy $h\nu_m$
of minimum-energy electrons \cite{Sari:1997qe}.  Here $\epsilon_e =
0.1\eps_{e,-1}$ is the fraction of shock energy converting to
accelerated electrons in the blastwave. Another break appears in the
opacity curve due to the characteristic synchrotron photon energy
$h\nu_c$ of cooling electrons and is given by
\be
E_{p,h} =  1.0\times 10^{12} 
\frac{\eps_{B,-1}^{3/2} n_0^{3/4} E_{55}^{3/4}}
{(1+z)^{3/4} t_2^{1/4}} ~{\rm GeV}.
\label{CRbreaks_ad_i_hi}
\ee
Note that $\tau_{p\g}$ and $E_{p,l}$ increase while $E_{p,s}$ and
$E_{p,h}$ decrease with time. These competing factors determine an
energy range $E_{p,l} < E_p <$ min\{$E_{p,s}$, $E_{p,h}$\} in which
$p\g$ interaction is the most efficient.

In the case of fast-cooling afterglow synchrotron spectrum
\cite{Sari:1997qe}, $h\nu_c < h\nu_m$, the $p\gamma$ opacity scales
with proton energy as
\ba
\tau_{p\gamma} (E_p) &=& \tau_{p\g} (E_{p,l}) \cr
&\times &
\begin{cases} 
\left( \frac{E_p}{E_{p,l}} \right)^{k/2} 
\,;\, E_p \lesssim E_{p,l} \cr
\left( \frac{E_p}{E_{p,l}} \right)^{1/2} 
\,;\, E_{p,l} \lesssim E_p \lesssim E_{p,h} \cr
\left( \frac{E_{p,h}}{E_{p,l}} \right)^{1/2} 
\,;\, E_p \gtrsim E_{p,h}, 
\end{cases}
\label{opacity_fast}
\ea
where $k=2.5$ is the electron spectrum typically used for synchrotron
afterglow modeling \cite{Sari:1997qe}, which is consistent with
modeling of X-ray afterglow data from Swift-XRT observations of a
large number of GRBs \cite{Racusin:2008bx}.  On the other hand, in the
case of slow-cooling afterglow synchrotron spectrum, $h\nu_c >
h\nu_m$, the $p\gamma$ opacity scales as
\ba
\tau_{p\gamma} (E_p) &=& \tau_{p\g} (E_{p,l}) \cr
&\times &
\begin{cases} 
\left( \frac{E_{p,h}}{E_{p,l}} \right)^{k/2-1/2} 
\left( \frac{E_p}{E_{p,h}} \right)^{k/2} 
\,;\, E_p \lesssim E_{p,h} \cr 
\left( \frac{E_p}{E_{p,l}} \right)^{k/2-1/2} 
\,;\, E_{p,h} \lesssim E_p \lesssim E_{p,l} \cr 
1  \,;\, E_p \gtrsim E_{p,l}.
\end{cases}
\label{opacity_slow}
\ea

The transition from fast-cooling to slow-cooling spectra happens when
$h\nu_c = h\nu_m$ or $E_{p,h} \approx E_{p,l}$ at a time
\be
t_{0} = 1.1\times 10^7\, (1+z) \eps_{B,-1}^2 \eps_{e,-1}^2 n_0 
E_{55} ~{\rm s}.
\label{t0_ad_ism}
\ee

To calculate neutrino fluxes from $p\gamma$ interactions, we calculate
an intermediate charged pion flux given by
\ba
J_\pi (E_\pi) &\approx& \frac{1}{2\langle x\rangle} 
J_p \left(\frac{E_\pi}{\langle x\rangle}\right)
\cr &\times& {\rm min}
\left\{ 
\tau_{p\g} \left(\frac{E_\pi (1+z)}{\langle x\rangle\Gamma}\right), 3
\right\},
\label{pi_flux}
\ea
where $\langle x\rangle \approx 0.2$ is the mean inelasticity for
$p\g\to\pi$ production through delta resonance. Finally, the neutrino
fluxes from pion decay are calculated by integrating over the product
of pion flux and various scaling functions for the $\pi^+ \to \mu^+
\nu_\mu \to e^+ \nu_e {\bar \nu}_\mu \nu_\mu$ chain decay
\cite{Lipari:1993hd}. As for example, $\pi^+$ decay $\nu_\mu$ flux is
\be
J_{\nu_\mu} (E_\nu) = \int_0^1 \frac{dx}{x} 
\frac{\Theta (1 - r_\pi - x)}{1 - r_\pi}
J_\pi \left( \frac{E_\nu}{x} \right)
\label{pion_nu_mu flux}
\ee
where $x = E_\nu/E_\pi$, $r_\pi = m_\mu^2/m_\pi^2$ and $\Theta$ is a
Heaviside step function. More details can be found in
Ref.~\cite{Razzaque:2013dsa}.

Neutrino fluxes from GRB blastwave last for a very long
time~\cite{Razzaque:2013dsa}, essentially until all kinetic energy is
dissipated. The intensity, however, progressively decreases over time.
We calculate $\nu$ fluxes until a time when the blastwave essentially
becomes non-relativistic with a Lorentz factor
\be
\Gamma \sim 1\,
\frac{(1+z)^{3/8} E_{55}^{1/8}}{n_0^{1/8} t_{7.5}^{3/8}}
\label{bw_G_ad_i}
\ee
after one year time scale, $t = 10^{7.5}t_{7.5}$ s. Note that the
dependence of this time scale on $E_k$ and $n_0$ is rather mild.  For
the Lorentz factor $\Gamma \gtrsim \theta_{\rm jet}^{-1} \approx 14
(\theta_{\rm jet}/0.06)$, where $\theta_{\rm jet}$ is the GRB jet
opening angle, the photon density in the blastwave decreases rapidly
at late time \cite{Sari:1999mr, Frail:2001qp}. We found, however, no
significant change in the time-intgrated flux or fluence as the
fluence is dominated by flux at earlier time (see
Ref.~\cite{Razzaque:2013dsa} and Fig.~2 therein).

\begin{figure}[t]
\includegraphics[width=2.95in]{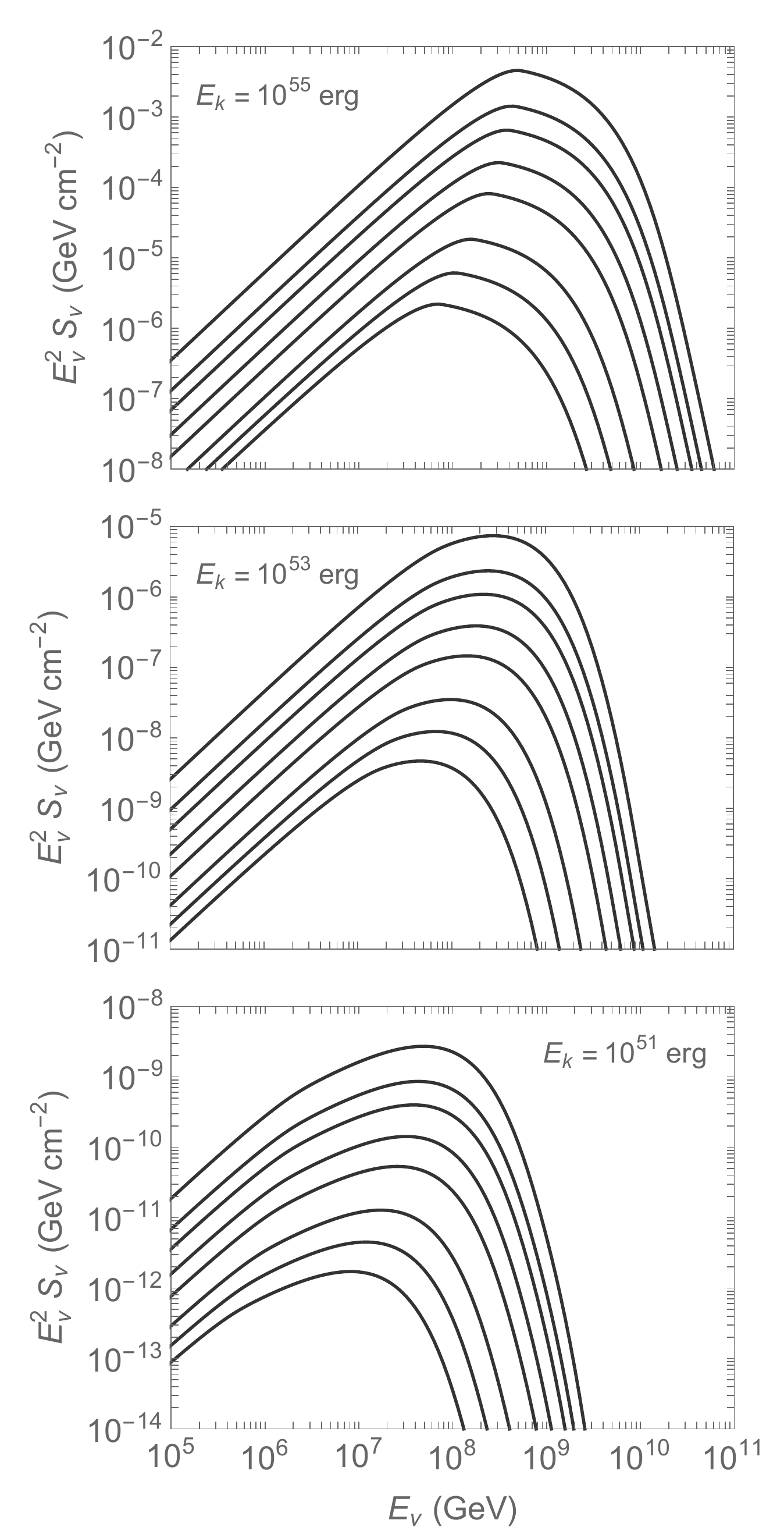}
\caption{Neutrino energy fluence ($\nu_\mu + {\bar \nu}_\mu$ after
  oscillation over astrophysical distances) from GRB blastwaves at
  redshift $z = 0.3,\, 0.5,\, 0.7,\, 1.1,\, 1.7,\, 3.3,\, 5.5,\, 9.0$
  from top to bottom curves in each panel. The top, middle and bottom
  panels correspond to $E_k = 10^{55}$ erg, $10^{53}$ erg and
  $10^{51}$ erg of isotropic-equivalent kinetic energy, respectively,
  of the blastwave in an interstellar medium of constant particle
  density $n_0 =1$~cm$^{-3}$.}
\label{fig:fluences}
\end{figure}

Figure \ref{fig:fluences} shows time integrated $\nu_\mu + {\bar
  \nu}_\mu$ energy flux (after oscillation over astrophysical
distance) or fluence $E_\nu^2 S_\nu$ evaluated for different blastwave
kinetic energies, $E_k = 10^{51}$--$10^{55}$ erg, (in different
panels) and for different redshift, $z = 0.3$--9.0, (as different
lines in each panel). We have assumed that $t_{dec} = 10$ s in
Eq.~(\ref{dec_time_ism}) is fixed for all GRBs and is determined by
$\Gamma_0$ as we have kept all other parameters fixed. The maximum
integration time is determined by Eq.~(\ref{bw_G_ad_i}). Note that the
$E_\nu$ at which the fluence curves peak is primarily determined by
the proton break energy $E_{p,l}$ in Eq.~(\ref{CRbreaks_ad_i_lo}),
above which $p\g$ opacity becomes significant and the maximum proton
energy $E_{p,s}$ in Eq.~(\ref{Emax_ad_i}), above which the proton
spectrum drops exponentially.

\subsection{Diffuse neutrino flux}

We calculate diffuse GRB blastwave neutrino flux by integrating
fluence of individual bursts over the kinetic energy and redshift
distributions of the observed rate of long-duration GRBs. We assume
that the isotropic-equivalent kinetic energy is given by
\be
E_k = \eta tL_\gamma
\label{EkLum}
\ee
where $L_\gamma$ is the isotropic-equivalent $\g$-ray luminosity of
long GRBs, $t \sim 10$ s is the typical duration of long GRBs and
$\eta$ is a dimensionless baryon-loading factor. We take $\eta t = 40$
s in our calculation as a parameter, following evidences for $\sim
20\%$ radiation efficiency for all GRBs \cite{Frail:2001qp} and
typical $\sim 10$ s duration of long GRBs.

Given a rate of long duration GRBs, ${\dot n}_{\rm GRB} (z, L)$ per
unit comoving volume element and per unit luminosity interval, the
diffuse neutrino flux can be derived as \cite{Razzaque:2005bh}
\ba
J_{\nu} (E_\nu) &=& \frac{1}{16\pi^2} \int_{z_1}^{z_2} \frac{dz}{1+z}
\frac{dV_c}{dz} \cr &\times&
\int_{L_1}^{L_2} dL \,{\dot n}_{\rm GRB} (z, L) S_\nu (E_\nu, z) 
\label{diff_flux}
\ea 
where the comoving volume element is given by
$$
\frac{dV_c}{dz} = \frac{4\pi c}{1+z} 
\left| \frac{dt}{dz} \right| d_L^2,
$$
with cosmic time and redshift relation
$$
\frac{dt}{dz} = \frac{-1}
{H_0 (1+z) \sqrt{\Omega_m (1+z)^3 + \Omega_\Lambda}}
$$
and luminosity distance
$$
d_L = c(1+z) \int_0^z dz^\prime (1+z^\prime) \frac{dt}{dz^\prime}.
$$
We use $H_0 = 69.6$ km s$^{-1}$ Mpc$^{-1}$, $\Omega_M = 0.286$ and
$\Omega_\Lambda = 0.714$ from the latest Planck results
\cite{Ade:2013zuv}. The upper and lower redshift values in
Eq.~(\ref{diff_flux}) are set as $z_2 = 9.0$ and $z_1 = 0.1$, based on
observations of long GRBs. The luminosity range is also set from
observations as $L_2 = 2.5\times 10^{53}$ erg/s and $L_1 = 2.5\times
10^{49}$ erg/s.

A recent fit to the {\it Swift} GRB data with redshift information
resulted in a luminosity function for long GRBs given by
\cite{Wanderman:2009es}
\begin{equation}
\phi(L) = \left\{
\begin{array}{l l} 
\left( \frac{L}{L_\star} \right)^{-0.17}; & L < L_\star \\
\left( \frac{L}{L_\star} \right)^{-1.44}; & L \ge L_\star ,
\end{array}  
\right.
\end{equation}
where $L_\star = 10^{52.53}$ erg/s. The corresponding redshift
evolution of GRB rate per unit comoving volume is
\begin{equation}
R(z)= R_0 \left\{
\begin{array}{l l} 
(1+z)^{2.07}; & z \le 3.11 \\
(1+z)^{-1.36} 4.11^{3.43}; & z > 3.11
\end{array}  
\right.
\end{equation}
where $R_0=1.25$ Gpc$^{-3}$ yr$^{-1}$ is the local GRB rate density.
Therefore,
\be
{\dot n}_{\rm GRB} (z, L) = R(z) \phi(L).
\label{GRB_rate}
\ee

\begin{figure}[t]
\includegraphics[width=3.25in]{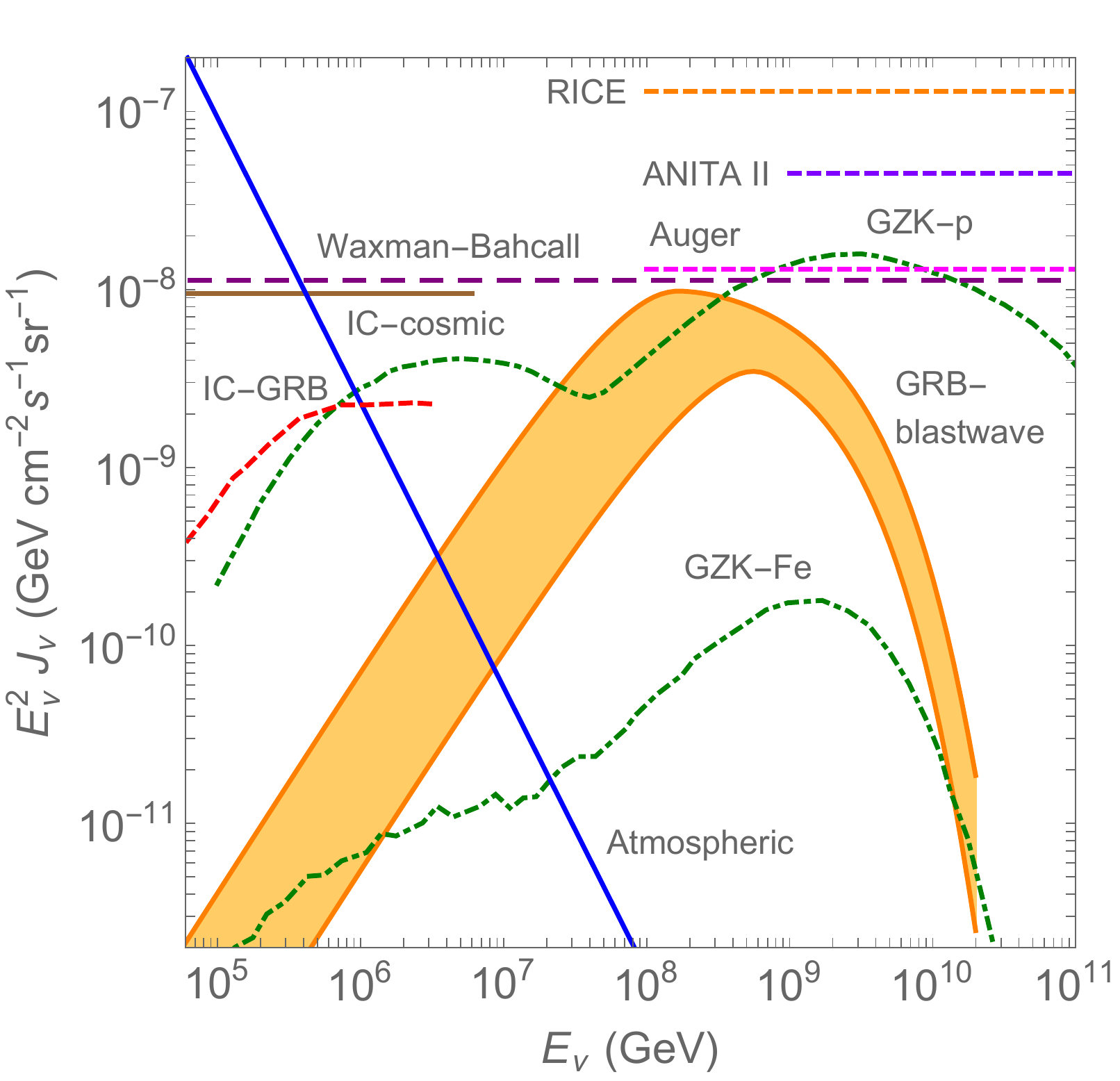}
\caption{Models of diffuse $\nu_\mu + {\bar \nu}_\mu$ fluxes from the
  GRB blastwave (shaded orange), atmosphere (solid blue)
  \cite{Honda:2006qj}, GZK $p$-dominated (upper dot-dashed green) and
  GZK Fe-dominated (lower dot-dashed green) \cite{Kotera:2010yn}.
  Also shown are the best-fit flux for the IceCube detected events
  (solid brown) \cite{Aartsen:2013jdh}, Waxman-Bahcall theoretical
  upper limit on the GZK flux \cite{Waxman:1998yy,Aartsen:2013jdh},
  IceCube upper limit on the prompt GRB flux \cite{Abbasi:2012zw}.
  Upper limits on diffuse $\propto E^{-2}$ flux from the Pierre Auger
  Observatory \cite{auger-nu}, ANITA-II \cite{Gorham:2010kv} and RICE
  \cite{Kravchenko:2011im} experiments.}
\label{fig:diffflux}
\end{figure}

Figure \ref{fig:diffflux} shows diffuse GRB blastwave $\nu_\mu + {\bar
  \nu}_\mu$ flux after oscillation over astrophysical distances
(shaded orange band) for two different circumburst particle densities,
$n_0 = 1$ cm$^{-3}$ (bottom curve) and 10 cm$^{-3}$ (upper curve).  A
larger value of $n_0$ results in a lower proton break energy $E_{p,l}$
in Eq.~(\ref{CRbreaks_ad_i_lo}) as well as a larger $p\g$ opacity in
Eq.~(\ref{opt_ad_i}). As a result the flux for $n_0 = 10$ cm$^{-3}$ is
higher and peaks at a lower energy. Also shown in
Fig.~\ref{fig:diffflux} are average atmospheric neutrino flux using a
model in Ref.~\cite{Honda:2006qj}, recently detected IceCube cosmic
neutrino flux (labeled ``IC-cosmic'') \cite{Aartsen:2013jdh}, as well
as limits on diffuse flux from the Auger Observatory \cite{auger-nu},
ANITA-II \cite{Gorham:2010kv} and RICE \cite{Kravchenko:2011im}.
IceCube limits \cite{Abbasi:2012zw} on prompt GRB neutrino flux is
labeled as ``IC-GRB'' while ``Waxman-Bahcall'' limit is based on
observed UHECR flux \cite{Waxman:1998yy,Aartsen:2013jdh}. We also show
cosmogenic neutrino flux models in Ref.~\cite{Kotera:2010yn}, denoted
as ``GZK-$p$'' and ``GZK-Fe'' in case UHECR primaries are proton and
iron, respectively.

\section{Detection rates}

Detection of PeV--EeV neutrinos from GRBs could be possible by the
IceCube Neutrino Observatory, the largest operating neutrino
telescope, by the Pierre Auger Observatory, the largest cosmic ray
detector, and by future facilities such as the high-energy extension
of IceCube, called IceHEX \cite{icehex}, the Askary'an Radio Array
(ARA) \cite{Allison:2011wk} and ARIANNA \cite{Barwick:2006tg}.  Here
we discuss prospects for detection of EeV neutrinos from individual
GRBs at low redshift and from a diffuse flux from GRBs in the history
of the universe.

\subsection{Individual GRBs}

\begin{figure}[t]
\vskip 0.12in
\includegraphics[width=3.15in]{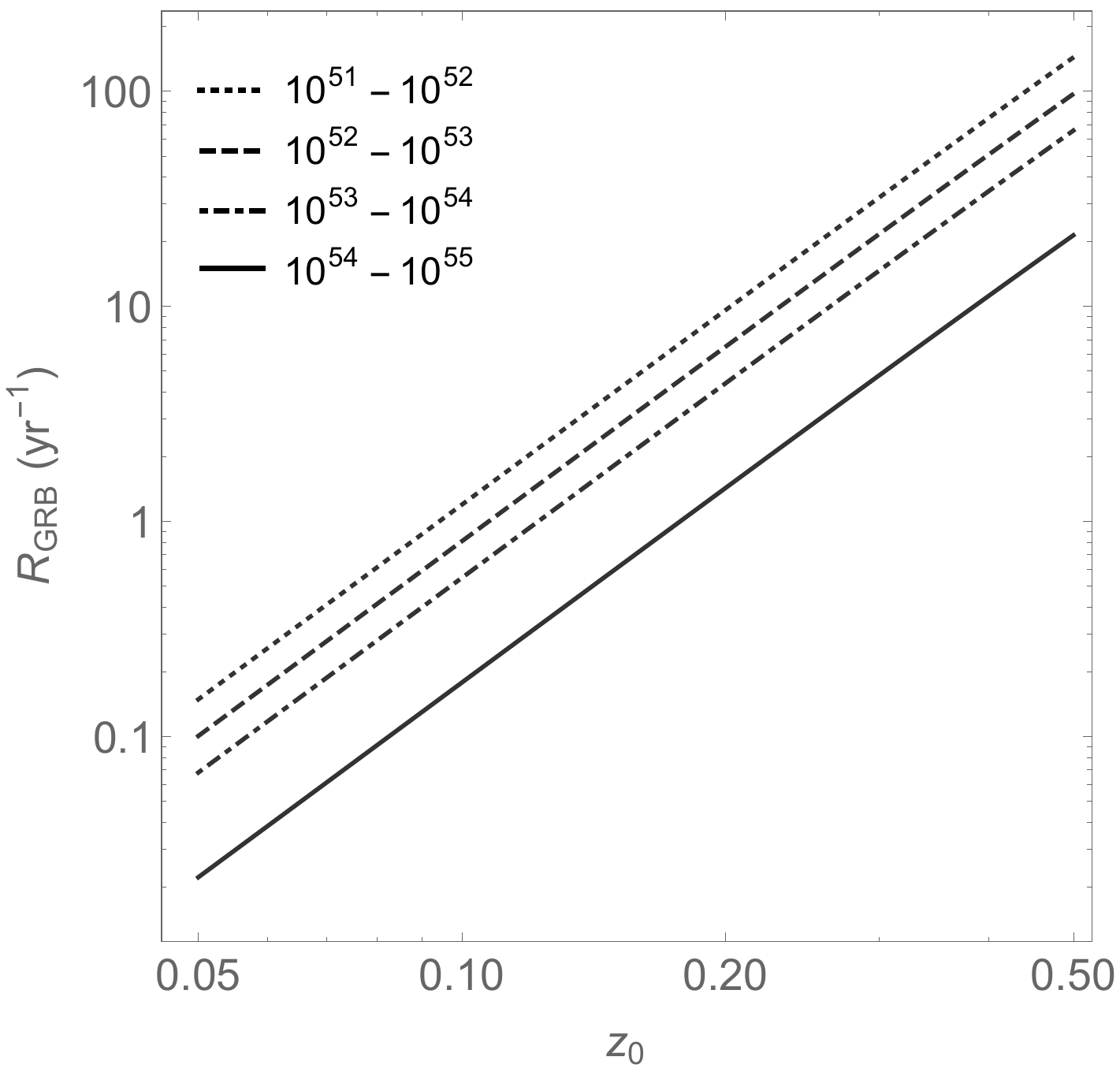}
\caption{The cumulative GRB rate [Eq.~(\ref{grb_z_rate})] as a
  function of redshift $z_0$ for different intervals of
  isotropic-equivalent blastwave kinetic energy $E_k$ in erg.}
\label{fig:grbrate}
\end{figure}

Neutrinos from individual GRBs at low redshift and high $\gamma$-ray
luminosity, which have higher fluxes at the Earth, could be detected.
The expected number of GRBs within redshift $z_0$ and within a
luminosity interval $L_1 \le L \le L_2$ can be calculated from the
rate in Eq.~(\ref{GRB_rate}) as,
\begin{equation}
R_{\rm GRB} (z_0)=\int_0^{z_0} \frac{dz}{1+z} \frac{dV_c}{dz}
\int_{\log L_1}^{\log L_2} d\log L \, {\dot n}_{\rm GRB} (z, L).
\label{grb_z_rate} 
\end{equation}
Figure~\ref{fig:grbrate} shows this rate as a function of redshift for
different intervals of $E_k$, the isotropic-equivalent blastwave
kinetic energy, related to the GRB isotropic-equivalent $\gamma$-ray
luminosity $L$ according to Eq.~(\ref{EkLum}). For example, the rate
of GRBs with $E_k$ in the $10^{54}$--$10^{55}$~erg is 1~yr$^{-1}$
within $z_0 \approx 0.2$.

Note that the Earth is essentially opaque to EeV neutrinos crossing a
significant fraction of its diameter. As a result, neutrino telescopes
are sensitive to downgoing neutrinos at EeV energies which reduces the
observable GRB rate by approximately half.

Individual GRBs appear as point sources to the neutrino telescopes.
The number of neutrino events of a particular neutrino flavor $\alpha$
from individual GRBs can be calculated as
\begin{equation}
N_{{\rm evt},\alpha} =  \int_{E_{\nu 1}}^{E_{\nu 2}} dE_\nu\, 
A_{{\rm eff},\alpha} (E_\nu) S^{\rm GRB}_{\nu_\alpha} (E_\nu),
\label{individual_sig}
\end{equation}
where $A_{{\rm eff},\alpha}$ is the effective area of the detector.
In principle, $A_{{\rm eff},\alpha}$ also depends on the zenith angle
from the detector. Here we take an average value over $4\pi$ solid
angle for IceCube \cite{Aartsen:2013bka}.  We chose the width of the
energy bins by considering the energy resolution of $\sim 0.3$-0.4 and
0.18 in $\log(E/{\rm GeV})$ scale for track like and cascade like
events, respectively, in IceCube \cite{Resconi:2008fe}.

At EeV energies, the background essentially is the expected diffuse
GZK neutrinos. The rate of these background events can be estimated as
\begin{equation}
N_{{\rm bkg},\alpha} =  2\pi \frac{\delta\theta^2}{2} T 
\int_{E_{\nu 1}}^{E_{\nu 2}} dE_\nu\, 
A_{{\rm eff},\alpha} (E_\nu) J^{\rm GZK}_{\nu_\alpha} (E_\nu),
\label{individual_bkg}
\end{equation}
where $\delta\theta$ is the angular resolution of the detector,
assumed to be small, and $T$ is the exposure time related to the GRB
flux duration time.

\begin{figure}[t]
  \includegraphics[width=3.in]{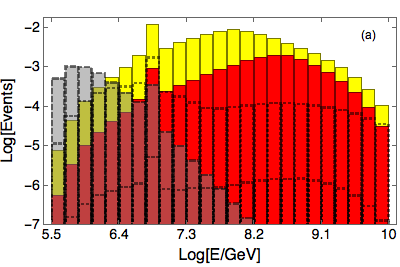}
  \includegraphics[width=3.in]{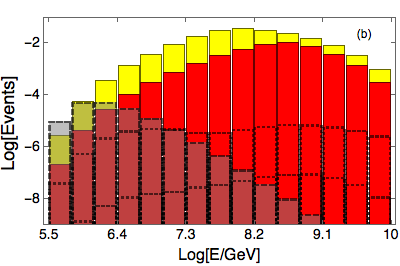}
  \includegraphics[width=3.in]{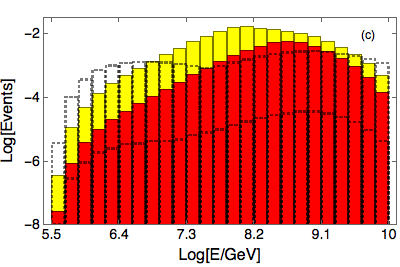}
  \caption{Expected energy distributions of neutrino and antineutrino
    events in IceCube from a GRB at redshift 0.2 with an interstellar
    medium of constant particle density 1~cm$^{-3}$ (red, solid edge,
    lower) and 10~cm$^{-3}$ (yellow, solid edge, higher). Also shown
    are the background event distributions from atmospheric (shaded,
    dashed edge), GZK-$p$ dominated (transparent, dotted edge, higher)
    and GZK-Fe dominated (transparent, dotted edge, lower) fluxes.
    The top, middle and bottom panels are for $e$-, $\mu$- and
    $\tau$-flavors respectively.}
  \label{fig:nu10r02}
\end{figure}

We calculate the expected neutrino events from individual GRBs and
backgrounds using IceCube effective areas~\cite{Aartsen:2013bka} and
$\delta\theta \sim 1^\circ$ angular resolutions for $\nu_\mu$ and
$\delta\theta \sim 10^\circ$ angular resolutions for $\nu_e$ or
$\nu_\tau$. For definiteness we place the GRB at $z=0.2$ with $E_k =
10^{55}$ erg. The rate of such GRBs is $\sim 1$~yr$^{-1}$ (see
Fig.~\ref{fig:grbrate}).  Figure \ref{fig:nu10r02} shows $\nu_e$,
$\nu_\mu$ and $\nu_\tau$ event distributions, both for the cases of
1~cm$^{-3}$ and 10~cm$^{-3}$ density of GRB circumburst medium. We
also show the distributions of background events in
Fig.~\ref{fig:nu10r02}.  These background events have been calculated
using atmospheric \cite{Honda:2006qj}, and GZK-$p$ and GZK-Fe flux
models \cite{Kotera:2010yn}. As can be seen, the background event
rates are extremely low. The enhancement of $\nu_e$ events at $\sim
6.3$ PeV is due to enhancement of the effective area for Glashow
resonance in the ${\bar \nu}_e$ cross section. Note that we have used
an exposure time of 1~yr, a time scale when the GRB blastwave becomes
non-relativistic [Eq.~(\ref{dec_time_ism})].

\begin{table}
  \caption{The expected number of $\nu_{\mu}$ events from a 
    GRB at $z=0.2$ and from backgrounds as plotted in Fig.~\ref{fig:nu10r02}.}
  \begin{center}
    \begin{tabular}{|c|c|c|c|}
      \hline
      $\log (E/{\rm GeV})$ & 5.5-7 & 7-8.5 & 8.5-10 \\
      \hline
      GRB 1/cm$^{-3}$ & $4.29\times 10^{-4}$ & $2.00\times 10^{-2}$ & 
      $2.33\times 10^{-2}$\\
      \hline         
      GRB 10/cm$^{-3}$ & $5.47\times 10^{-3}$ & $0.12$ & 
      $4.88\times 10^{-2}$\\
      \hline         
      Atmospheric & $1.47 \times 10^{-4}$ & $6.36\times10^{-6}$ & 
      $1.20 \times 10^{-8}$\\
      \hline        
      GZK-Fe & $3.24\times 10^{-8}$ & $1.92\times 10^{-7}$ & 
      $2.71\times 10^{-7}$ \\
      \hline
      GZK-$p$ & $1.11\times 10^{-5}$ & $2.12\times 10^{-5}$ & 
      $2.52\times 10^{-5} $ \\                 
      \hline
    \end{tabular}
  \end{center}
  \label{tabrate1}
\end{table}

Table \ref{tabrate1} shows $\nu_\mu$ track events in
Fig.~\ref{fig:nu10r02} for larger energy bins. As expected from flux
models, the number of events is larger for 10~cm$^{-3}$ density.  Note
that the prospect for detection of individual GRBs by IceCube is
better for a GRB at $z < 0.2$ and with $E_k > 10^{55}$ erg. The rate
of such GRBs is 1 per $\sim 5$--10 yr (see Fig.~\ref{fig:grbrate}). We
will comment on the detectability of individual GRBs by a future
high-energy extension of the IceCube, called IceHEX \cite{icehex}, in
Sec.~\ref{subsec:future}.

\subsection{Neutrinos from diffuse flux}

The number of $\nu$ events from diffuse GRB or background fluxes,
$J^{\rm diff}_{\nu_\alpha}$, can be calculated as
\begin{equation}
N_{{\rm evt},\alpha} =  4\pi T \int_{E_{\nu 1}}^{E_{\nu 2}} dE_\nu\, 
A_{{\rm eff},\alpha} (E_\nu) J^{\rm diff}_{\nu_\alpha} (E_\nu),
\label{diffuse_sig}
\end{equation}
where $A_{{\rm eff},\alpha}$ is the average effective area over $4\pi$
solid angle \cite{Aartsen:2013bka} as noted before and $T$ is the
exposure time. 

\begin{figure}[t]
  \includegraphics[width=3.in]{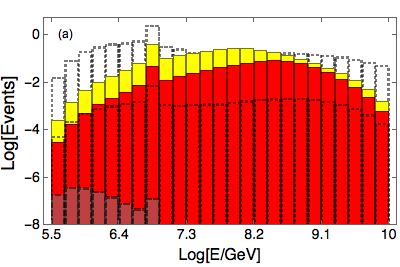}
  \includegraphics[width=3.in]{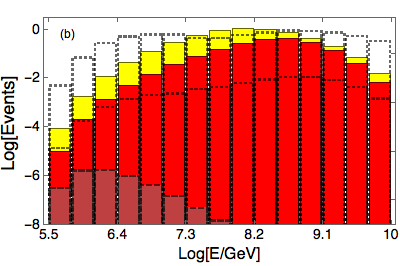}
  \includegraphics[width=3.in]{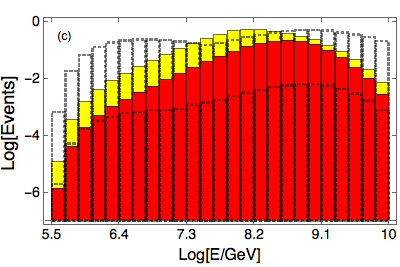}
  \caption{Expected energy distributions of diffuse neutrino and
    antineutrino events in IceCube from GRBs with an interstellar
    medium of constant particle density 1 cm$^{-3}$ (red, solid edge,
    lower) and 10 cm$^{-3}$ (yellow, solid edge, higher).
    Distributions for atmospheric (shaded, dashed edge), GZK-$p$
    dominated (transparent, dotted edge, higher) and GZK-Fe dominated
    (transparent, dotted edge, lower) flux models are also shown.
    Events are calculated for 10 years of IceCube running time for
    $e$- (top panel), $\mu$- (middle panel) and $\tau$- (bottom panel)
    flavors.}
  \label{fig:nu10}
\end{figure}

Figure \ref{fig:nu10} shows energy distributions of neutrino events in
IceCube for different flavors from diffuse GRB neutrino fluxes and
from backgrounds plotted in Fig.~\ref{fig:diffflux}. We have used 10
yr exposure time for this calculation. While the background events
from GZK-$p$ flux \cite{Kotera:2010yn} dominate most of the energy
range, events from difuse GRB flux are above all backgrounds in the
$\sim 10^{7.5}$-$10^{8}$ GeV range in case of $10$ cm$^{-3}$ density
of the GRB circumburst medium.  Table \ref{tabrate2} lists the number
of events in larger energy bins and for 5 year IceCube exposure.  Note
that detection of diffuse GRB blastwave neutrino flux is most probable
in case the UHECR primaries are Fe, as suggested by the Pierre Auger
Observatory \cite{Abraham:2010yv}. In such a case UHE protons
accelerated in the GRB blastwaves will not contribute significantly to
the observed UHECR flux.

The Pierre Auger Observatory is sensitive to EeV neutrinos. The
current limit on $E^{-2}$ diffuse flux from the Pierre Auger
Observatory \cite{auger-nu} is the most stringent in the 0.1--100 EeV
range, however it does not constrain the diffuse GRB blastwave flux
models presented in this work (see Fig.~\ref{fig:diffflux}).

\begin{table}
  \caption{The expected event numbers for $\nu_e$, $\nu_{\mu}$ 
    (in parentheses) and $\nu_{\tau}$ (in brackets) from diffuse GRB 
    fluxes and backgrounds in Fig.~\ref{fig:diffflux} for a 5-year 
    IceCube search over the full sky.}
  \begin{center}
    \begin{tabular}{|c|c|c|c|}
      \hline
      $\log(E/{\rm GeV})$ & 5.5--7 & 7--8.5 & 8.5--10 \\
      \hline
      & 0.03 & 0.18 & 0.13 \\
      GRB 1/cm$^{-3}$ & (0.01) & (0.44) & (0.44) \\
      & [0.01] & [0.33] & [0.43]\\
      \hline
      & 0.26 & 0.83 & 0.21 \\
      GRB 10/cm$^{-3}$ & (0.09) & (1.91) & (0.71) \\
      & [0.06] & [1.40] & [0.69] \\
      \hline
      & $7.33 \times 10^{-7}$ & $5.14\times 10^{-9}$ & 
      $4.29\times 10^{-12}$ \\
      Atmospheric & ($2.50 \times 10^{-6}$) & ($1.08\times10^{-7}$) & 
      ($2.04 \times 10^{-10}$) \\
      & [---] & [---] & [---] \\
      \hline
      & $6.12\times 10^{-3}$ & $5.44\times 10^{-3}$ & 
      $5.00\times 10^{-3}$ \\
      GZK-Fe & ($2.12\times 10^{-3} $) & ($1.26\times 10^{-2}$) & 
      ($1.78\times 10^{-2}$) \\
      & [$1.65 \times 10^{-3} $] & [$9.33\times 10^{-3}$] & 
      [$1.79\times 10^{-2}$] \\
      \hline
      & 2.11 & 0.67 & 0.46 \\
      GZK-${p}$ & (0.73) & (1.39) & (1.66) \\
      & [0.58] &[0.99] & [1.70] \\
      \hline
    \end{tabular}
  \end{center}
  \label{tabrate2}
\end{table}

\subsection{Detectability by Future Neutrino Telescopes}
\label{subsec:future}

There are several large, with $\gtrsim 100$ km$^2$ geometric area,
neutrino telescopes currently at the proposal stage \cite{icehex,
  Allison:2011wk, Barwick:2006tg}. The threshold energy for these
detectors is expected to be in the $\sim 10$-100 PeV range. The
prospect for GRB blastwave neutrino detection by these large scale
neutrino telescopes is very promising. Here we estimate event numbers
for the proposed high-energy extension of IceCube \cite{icehex}, which
we refer to as IceHEX, by extrapolating some characteristics of
IceCube.

The effective area for a detector in case of downgoing neutrinos can
be calculated, after taking into account $\nu$ survival probability
and $\nu N$ interaction probability within the detector, as
\cite{Razzaque:2003uw, Ioka:2005er}
\be 
A_{\rm eff} (E_\nu) = 7\times 10^{-5} \epsilon_{\rm det} 
A_{\rm geo} \left( \frac{E_\nu}{10^{4.5}~{\rm GeV}} \right)^\beta, 
\label{Aeffsimple}
\ee
where $\beta = 1.35$ for $E_\nu < 10^{4.5}$ GeV while $\beta = 0.55$
for $E_\nu \ge 10^{4.5}$ GeV, $A_{\rm geo}$ is the geometric area of
the detector and $\epsilon_{\rm det} < 1$ is a detector efficiency
factor which is energy dependent in general. This parameterization
reasonably reproduces the IceCube effective area with $A_{\rm geo} =
1$ km$^2$ and $\epsilon_{\rm det} \sim 0.1$ for $E_\nu \gtrsim 10^7$
GeV and for cascade events.

For IceHEX, we take the proposed geometric area $A_{\rm geo} = 100$
km$^2$ and fix $\epsilon_{\rm det} = 0.1$, the same as IceCube.  For a
GRB blastwave at $z=0.2$, the expected number of cascade ($\nu_e$ and
$\nu_\tau$) events are 10 and 40 for GRB circumburst density
1~cm$^{-3}$ and 10~cm$^{-3}$, respectively, for $10^7 < E_\nu <
10^{10}$ GeV.  Therefore neutrinos can be detected from a GRB
blastwave up to a redshift $z\sim 0.5$. IceHEX will be able to detect
diffuse flux of neutrinos from GRB blastwave as modeled in this work,
at a rate of 20--100 cascade events/yr, depending on the flux level.

\section{Discussion and conclusions} 

We have calculated neutrino fluence from GRB blastwaves in the
PeV--EeV range, following Ref.~\cite{Razzaque:2013dsa} for individual
GRBs (See Fig.~\ref{fig:fluences}).  The detailed diffuse flux
calculation is based on the observed rate of long GRBs and their
redshift evolution.  The diffuse flux per neutrino flavor peaks in the
0.1--1 EeV range (see Fig.~\ref{fig:diffflux}) and is lower than the
diffuse cosmic neutrino flux detected by IceCube
\cite{Aartsen:2013jdh}, below 2 PeV. Our optimistic model of GRB
blastwave neutrino flux peaks slightly below the Waxman-Bahcall limit
\cite{Waxman:1998yy, Aartsen:2013jdh} and the conservative one is a
factor 5 lower. Cosmogenic neutrino flux modeled in
Ref.~\cite{Kotera:2010yn} with UHECR proton primary is higher than our
model of GRB blastwave flux, except for our optimistic flux model and
in the 0.03--0.3 EeV range. We do not require protons accelerated in
the GRB blastwave to contribute to the observed UHECRs dominantly.
Therefore if UHECRs are primarily heavy (iron) nuclei, our model of
flux can dominantly contribute to the diffuse neutrino background in
the PeV--EeV range.

We have also calculated neutrino events, using our diffuse flux model,
in IceCube and future large neutrino telescopes referred to as IceHEX
here, which is a planned high-energy upgrade of IceCube \cite{icehex}.
We find that IceCube can detect neutrinos from GRB blastwave after 5
years of operation if our diffuse flux model is correct. This would
constitute detection of this flux component, in case UHECR primaries
are heavy nuclei.  In case UHECR primaries are protons, identification
of the GRB blastwave diffuse flux component will take a longer time.
With a 100 times larger effective area than IceCube at EeV energies,
IceHEX can detect 20-100 cascade events/yr from GRBs if our diffuse
flux model is valid.

Detection of neutrinos by current IceCube from an individual GRB
blastwave may not be possible. Stacking analysis with individual GRBs,
however, can improve the sensitivity. An observation strategy with
long-term ($\sim 1$ year time scale) monitoring of GRBs is needed.
IceHEX will be able to detect 10's of events from an energetic GRB at
$z=0.2$ for which the rate is 1/yr (see Fig.~\ref{fig:grbrate}). Since
the neutrino events from individual GRBs are essentially background
free, few events may constitute a detection and IceHEX may detect GRBs
at redshift up to $z\sim 0.5$, where the rate is roughly 10 times
higher.  Such a detection will be crucial to understand GRB explosion
energy, environment and most importantly acceleration of particles to
ultrahigh energies.

\acknowledgments

This work was supported in part by the National Research Foundation
(South Africa) grants nos.\ 87823 (CPRR) and 91802 (Blue Skies), to
SR. We thank Kohta Murase for useful comments, in particular about the
GRB jet break.

\end{document}